\documentclass[showpacs,preprintnumbers,amsmath,twocolumn,amssymb]{revtex4}
\usepackage{graphicx}
\usepackage{dcolumn}
\usepackage{bm}
\usepackage{amssymb}
\usepackage{array}
\usepackage{hhline}
\usepackage{longtable}

\begin{document}

\title{Degree and component size distributions in generalized uniform recursive tree}

\author{Zhongzhi Zhang$^{1,2}$}
\email{zhangzz@fudan.edu.cn}

\author{Shuigeng Zhou$^{1,2}$}
\email{sgzhou@fudan.edu.cn}

\author{Shanghong Zhao$^{1,2}$}

\author{Jihong Guan$^{3}$}

\author{Tao Zou$^{1,2}$}

\affiliation {$^{1}$Department of Computer Science and Engineering,
Fudan University, Shanghai 200433, China}

\affiliation {$^{2}$Shanghai Key Lab of Intelligent Information
Processing, Fudan University, Shanghai
200433, China} %

\affiliation{$^{3}$Department of Computer Science and Technology,
Tongji University, 4800 Cao'an Road, Shanghai 201804, China}


\begin{abstract}
We propose a generalized model for uniform recursive tree (URT) by
introducing an imperfect growth process, which may generate
disconnected components (clusters). The model undergoes an
interesting phase transition from a singly connected network to a
graph consisting of fully isolated nodes. We investigate the
distributions of degree and component sizes by both theoretical
predictions and numerical simulations. For the nontrivial cases, we
show that the network has an exponential degree distribution while
its component size distribution follows a power law, both of which
are related to the imperfect growth process. We also predict the
growth dynamics of the individual components. All analytical
solutions are successfully contrasted with computer simulations.

\end{abstract}

\date{\today}
\maketitle

Degree distribution and component (cluster) size distribution are
two important properties~\cite{CoRoTrVi07} of complex networks,
which have become a focus of attention for the scientific
community~\cite{AlBa02,DoMe02,Ne03,BoLaMoChHw06}. It has been
established that degree distribution has an important consequence on
other structural characteristics of networks~\cite{CoRoTrVi07}, such
as average path length, clustering coefficient, betweenness
centrality, and among others. Particularly, degree distribution has
also profound effects on almost all aspects on dynamic processes
taking place on networks, including robustness~\cite{AlJeBa00},
percolation~\cite{CaNeStWa00,CoErAvHa01},
synchronization~\cite{BaPe02}, games~\cite{SzFa07}, epidemic
spreading~\cite{PaVe01}, and so on. On the other hand, as a
fundamental structural feature of networks, component (especially
the giant component) size is one of the most significant measures of
the network robustness~\cite{AlJeBa00,CaNeStWa00,CoErAvHa01}; in
particular, component size distribution is directly related to many
significant practical issues such as the distribution of the sizes
of disease outbreaks for disease propagation on contact
networks~\cite{Gr82,Ne02,Ne07}.

A wide variety of network models has been proposed to describe
real-life systems and study their structural
properties~\cite{AlBa02,DoMe02,Ne03,BoLaMoChHw06}, among which
uniform recursive tree (URT) is perhaps one of the earliest and most
widely studied simplest models. The URT is a growing tree
constructed as follows: Start with a single node, at each increment
of time, let a new node be added to the network linked to a randomly
selected preexisting node. It has found applications in several
areas. For example, it has been suggested as models for the spread
of epidemics~\cite{Mo74}, the family trees of preserved copies of
ancient or medieval texts~\cite{NaHe82}, chain letter and pyramid
schemes~\cite{Ga77}, to name but a few.

Most previous models belong to idealized
models~\cite{AlBa02,DoMe02,Ne03,BoLaMoChHw06}, although they may
provide valuable insight into reality. In fact, real-world systems
may be viewed as imperfect realizations of idealized theoretical
models~\cite{ZeKaSt07}, which have significant influences on their
features and function. For instance, introducing defects
(impurities) into magnetic materials may drastically change its
property~\cite{HuBe89}. Recently, many authors have focused on
investigating the influence of deleting nodes (or edges) on the
function of networks such as their integrity and communication
ability~\cite{AlJeBa00,CoErAvHa01,CrLaMaRa03}.

In this paper, we present a generalized model of URT by introducing
an imperfect growth process which may generate disconnected
components. The proposed model is governed by a tunable parameter
$q$. We study both analytically and numerically the interesting
characteristic aspects of the models, focusing on degree
distribution, component size distribution, as well as the evolving
dynamics of individual componets. We show that both distributions
can be computed accurately, although calculation of component size
distribution seems difficult for previously studied models. The
obtained results indicate that the imperfect growth process affects
fundamentally the structure of the network.

\label{sec:1} The proposed generalized model with an imperfection
growth process is generated in the following way. We start from an
initial state ($t=0$) of one isolated node. Then, at each increment
of time, a new node is added, whose behavior of growth dynamics
depends on a parameter $q$. With probability $q$, the new node keeps
isolated; and with complementary probability $1-q$, the newly
introduced node connects a randomly chosen old node. This growing
process is repeated until the network reaches the desired size. One
can easily see that at time $t$, the network consists of $t+1$ nodes
and expected $(1-q)t$ edges. Thus, when $t$ is large, the average
node degree at time $t$ is approximately equal to a constant value
$2(1-q)$.

There are two limiting cases of the present model with URT as one of
its particular case. Therefore, we call the presented model
``generalized uniform recursive tree". For $q=0$, the model
coincides with the uniform recursive tree which is a singly
connected network. When $q=1$, the network is reduced to a fully
disconnected network. Thus, varying $q$ in the interval (0,1) allows
the formation of disconnected clusters as part of the growing
dynamics.


Below we will show that some interesting characteristics (i.e.,
degree distribution, component size distribution and evolutionary
dynamics of individual components) may be investigated analytically,
which depend on the parameter $q$, i.e., the imperfect growth
process.

First we focus on the degree distribution. For $q=1$, all nodes have
the same number of connection 0, the network exhibits a completely
homogeneous degree distribution. For the case of $0\leq q <1$, we
can address the degree distribution using the master-equation
approach~\cite{DoMeSa00}. For simplicity, we label nodes by their
time of birth, so that node $s$ refers to the node introduced at
time $s$, and use $p(k,s,t)$ to denote the probability that at time
$t$ the node $s$ has a degree $k$. Then we may analyze the dynamics
through a set of master equations~\cite{DoMeSa00,DoMeSa01,ZhRoZh07}
governing the evolution of the degree distribution of an individual
node, which are of the following form
\begin{eqnarray}\label{eq1}
p(k,s,t+1)=\frac{1-q}{t+1}p(k&-&1,s,t)+\left(1-\frac{1-q}{t+1}\right)p(k,s,t)\nonumber\\
 &+&q\delta_{k,0}+(1-q)\delta_{k,1}
\end{eqnarray}
with the initial condition, $p(k,s=0,t=0)=\delta_{k,0}$. This
accounts for two possibilities for a node: first, with probability
$\frac{1-q}{t+1}$, it may get an extra edge from the new node, and
thus increase its own degree by 1; and second, with the
complimentary probability $1-\frac{1-q}{t+1}$, the node may remain
in the former state with the former degree. 

Based on Eq.~(\ref{eq1}), one can obtain the total degree
distribution $P_t(k)$ specifying the probability that a randomly
chosen node has degree $k$ at time $t$:
\begin{equation}\label{eq2}
P_t(k)=\frac{1}{t+1}\sum_{s=0}^{t}p(k,s,t).
\end{equation}
Summing up both sides of Eq.~(\ref{eq1}) over $s$, we get the
following master equation for the degree distribution:
\begin{eqnarray}\label{eq3}
&\quad&(t+2)P_{t+1}(k)-(t+1)P_t(k)\\\nonumber
&=&(1-q)P_t(k-1)-(1-q)P_t(k)+q\delta_{k,0}+(1-q)\delta_{k,1}.
\end{eqnarray}
In the infinite $t$ limit, each $P_t(k)$ converges to some lime
$P(k)$. Then, the corresponding stationary equation takes the form
\begin{equation}\label{eq4}
(2-q)P(k)-(1-q)P(k-1)=q\delta_{k,0}+(1-q)\delta_{k,1}.
\end{equation}
It has the solution of an exponential form
\begin{equation}\label{eq5}
P(k)=\left\{\begin{array}{lc} {\displaystyle{\frac{q}{2-q},} } & \ \
k=0,\\
{\displaystyle{\frac{2}{2-q}\left(\frac{1-q}{2-q}\right)^k,} }
& \ \ k\geq 1.\\
\end{array} \right.
\end{equation}
For $q=0$, Eq.~(\ref{eq5}) is exactly the same degree distribution
of the uniform recursive tree~\cite{DoMe02}.

\begin{figure}
\begin{center}
\includegraphics[width=8cm]{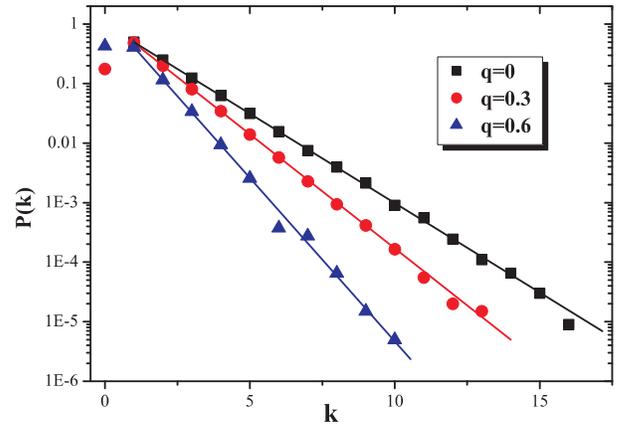}
\end{center}
\caption{(Color online) Semilogarithmic graph of degree distribution
of the networks with order $N=200000$. Each point is an average over
10 independent simulations. The solid lines are the analytic
calculation values given by Eq.~(\ref{eq5}).} \label{fig1}
\end{figure}

In Fig.~\ref{fig1}, we report the simulation results of the degree
distribution for several values of $q$, from which we can see that
the degree distribution decays exponentially with the degree, in
agreement with the analytical results.


Next, we show that the distribution of component sizes follows a
power law with exponent determined by parameter $q$. For the case of
$q=0$, there is only one component in the network, and the component
size distribution is trivial. For $q=1$, there are $t+1$ components,
every node belongs to a component of size 1 (the node itself). In
the range $0<q<1$, the imperfect growth process allows the formation
of multiple disconnected components. For this case, we can derive an
analytical expression for the component size distribution using the
rate-equation method~\cite{KaReLe00}.

Initially ($t=0$), there is only one component in the network. At
each subsequent step, a new cluster is created with probability $q$.
Thus, after $t$ step evolution, there are expected $N(c,t)=tq+1$
components in the network, and the average value of component sizes
is asymptotically equal to $\frac{1}{q}$ for large $t$. Let $N_c(t)$
denote the average number of components with size $c$ at time $t$.
By the very construction of the generalized uniform recursive tree,
when a new node enters the network, the rate
equations~\cite{KaReLe00,KaRe01,WaDuDaSu06,ZhZh07} that account for
the evolution of $N_c(t)$ with time $t$ are
\begin{equation}\label{rate}
\frac{dN_c(t)}{dt}=(1-q)\frac{(c-1)
N_{c-1}(t)-c\,N_c(t)}{{\sum_{c'}{c'\,N_{c'}(t)}}}+q\,\delta_{c,1}.
\end{equation}
Here the first term on the right-hand side of Eq.~(\ref{rate})
accounts for the process in which the new node is connected to one
node in a cluster with size $c-1$, leading to a gain in the number
of components with size $c$. Since there are $N_{c-1}(t)$ components
of size $c-1$ and the new node creates a new edge with probability
$1-q$, such processes occur at a rate proportional to $(1-q)(c-1)
N_{c-1}(t)$, while the factor ${\sum_{c'}{c'\,N_{c'}(t)}}$ converts
this rate into a normalized probability. The second (loss) term
describes the new edge connecting to one of the nodes in a cluster
with size $c$ turning it into a component with size $c+1$. The last
term on the right-hand side of Eq.~(\ref{rate}) accounts for the
continuous introduction of a new component with size 1 (the isolated
new node itself).

Let $P(c)$ be the component size distribution that is the
probability of a randomly chosen cluster having size $c$. In the
asymptotic limit $N_c(t)=N(c,t)P(c) \simeq tqP(c)$ and
${\sum_{c'}{c'\,N_{c'}(t)}}=t+1$. Inserting these into
Eq.~(\ref{rate}), we have the following recursive equation
\begin{equation}\label{eq7}
P(c)=\left\{\begin{array}{lc}
{\displaystyle{\frac{c-1}{c+\frac{1}{1-q}}P(c-1),} }
& \ \ c>1,\\
{\displaystyle{\frac{1}{2-q},} }
& \ \  c=1,\\
\end{array} \right.
\end{equation}
giving
\begin{equation}\label{eq8}
P(c)=\frac{1}{2-q}\prod_{m=2}^{c}\frac{m-1}{m+\frac{1}{1-q}}=\frac{1}{2-q}\frac{\Gamma
(c) \Gamma \left(2+\frac{1}{1-q}\right)}{\Gamma
\left(c+1+\frac{1}{1-q}\right)}
\end{equation}
for $c>1$. Thus, in the infinite $c$ limit
\begin{equation}\label{eq9}
P(c)\thicksim c^{-\left(1+\frac{1}{1-q}\right)},
\end{equation}
which shows that component size distribution follows a power-law
form with an exponent $\gamma_c=1+\frac{1}{1-q}$ dependent on
parameter $q$.

\begin{figure}
\begin{center}
\includegraphics[width=8cm]{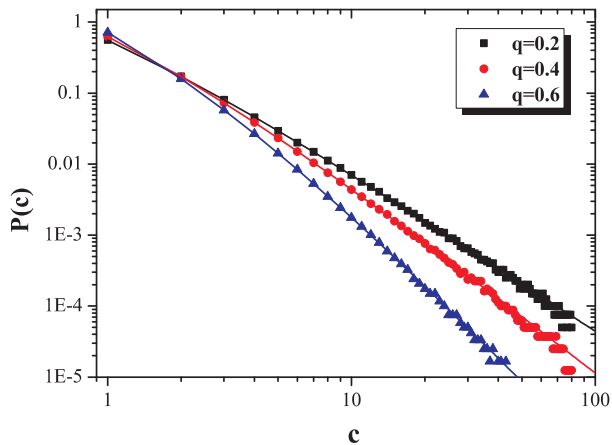}
\end{center}
\caption{(Color online) Double-logarithmic plot of the component
size distribution of the networks with 200000 nodes each. The points
are the results of computer simulations, and each point is obtained
by 20 independent network realizations. The solid lines correspond
to the analytic solution provided by Eq.~(\ref{eq8}).} \label{fig2}
\end{figure}

We have performed extensive numerical simulations for the full range
of $q$ between 0 and 1. 
In Fig.~\ref{fig2}, we plot the component size distribution $P(c)$
as a function of $q$, which agrees well with the analytic result.

\begin{figure}
\begin{center}
\includegraphics[width=8cm]{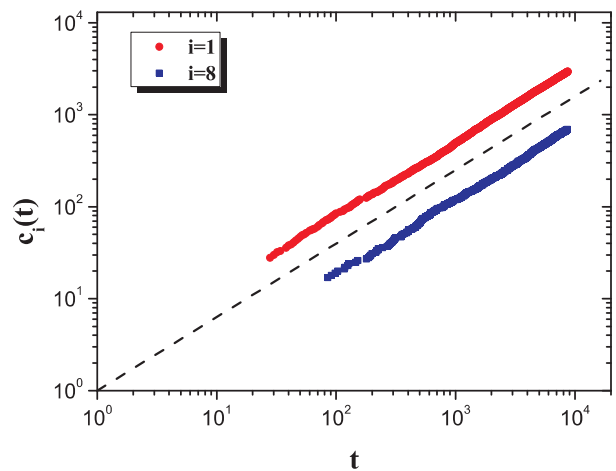}
\end{center}
\caption{(Color online) Double-logarithmic graph showing the
time-evolution for the sizes of two components. Here $q=0.2$, and
the dashed line has slope 0.8, as predicted by Eq.~(\ref{eq11}).}
\label{fig3}
\end{figure}

In addition to component size distribution, using the continuum
approach~\cite{BaAlJe99} we can calculate the time dependence of
component size $c_i$ of a given component $i$, $i=1,2,\cdots,
N(c,t)$. The size will increase by one every time a new node is
added to the system and linked to one of the nodes belonging to this
component. Assuming that $c_i$ is a continuous real variable,
according to the generating algorithm of the model, the rate at
which $c_i$ changes is clearly proportional to $c_i$ itself.
Consequently $c_i$ satisfies the dynamical
equation~\cite{BaAlJe99,ZhRoCo06}
\begin{equation}\label{eq10}
\frac{\partial c_i(t)}{\partial t} = (1-q)\frac{c_i(t)}{t}.
\end{equation}
The solution of this equation, with the initial condition that
component $i$ was born at time $t_i$ with size $c_i(t_i)=1$, is
\begin{equation}\label{eq11}
c_i(t)=\left(\frac {t}{t_i}\right)^{1-q}.
\end{equation}
Equation~(\ref{eq11}) shows that the size of all components evolves
the same way, following a power law, the only difference being the
intercept of this power law, see Fig.~\ref{fig3}.

To conclude, by introducing an imperfect network growth process we
have proposed an extended model for uniform recursive tree. The
presented model interpolates between a network with fully
disconnected clusters and the uniform recursive tree with only one
single component, which allow us to explore the crossover between
the two limiting cases. We have provided both analytically and
numerically the solutions for degree and component size
distributions of our model. We found that in the case of $0 < q < 1$
the model exhibits an exponential degree distribution and a
power-law component size distribution. We also presented that the
size of all components evolves as a power-law function of time $t$.

Our model has a remarkable character that some nodes may become
disconnected from the rest of the network. This property has been
less reported in earlier studied models and thus has not yet been
paid enough attention to. Actually, our model may be further
extended to include initial attractiveness~\cite{DoMeSa00} and
preferential attachment~\cite{BaAl99}. That is, all nodes are born
with some initial attractiveness $A>0$. In the growth progress, the
probability that an old node will receive an link from the new old
is proportional to the sum of the initial attractiveness and its
degree. In this more general case, the degree distribution is power
law with exponent dependent on $A$, while the component size
distribution is the same as that of the present model addressed here
and is independent of parameter $A$. Finally, we believe our model
could stimulate and find some obvious implications for some related
researches such as disease propagation in future.

This research was supported by the National Basic Research Program
of China under grant No. 2007CB310806, the National Natural Science
Foundation of China under Grant Nos. 60496327, 60573183, 90612007,
60773123, and 60704044, the Postdoctoral Science Foundation of China
under Grant No. 20060400162, the Program for New Century Excellent
Talents in University of China (NCET-06-0376), and the Huawei
Foundation of Science and Technology (YJCB2007031IN).

\end{document}